\documentclass[12pt,preprint]{aastex}

\def\gtorder{\mathrel{\raise.3ex\hbox{$>$}\mkern-14mu
             \lower0.6ex\hbox{$\sim$}}}

\def\ltsima{$\; \buildrel < \over \sim \;$}
\def\simlt{\lower.5ex\hbox{\ltsima}}
\def\gtsima{$\; \buildrel > \over \sim \;$}
\def\simgt{\lower.5ex\hbox{\gtsima}}

\begin{document}


\title{A Complete Survey of the Transient Radio Sky
and Implications for Gamma-Ray Bursts, Supernovae,
and other Relativistic Explosions}


\author{Avishay Gal-Yam\altaffilmark{1}}
\affil{Astronomy Department, MS 105-24, California Institute of Technology,
Pasadena, CA 91125}
\email{avishay@astro.caltech.edu}
\author{Eran O. Ofek, Dovi Poznanski, Amir Levinson}
\affil{School of Physics \& Astronomy, Tel-Aviv University, Tel-Aviv
69978, Israel}
\author{Eli Waxman}
\affil{Department of Condensed Matter Physics, Weizmann Institute, Rehovot 76100, Israel}
\author{Dale A. Frail}
\affil{National Radio Astronomy Observatory, P.O. Box O, Socorro, NM 87801}
\author{Alicia M. Soderberg, Ehud Nakar}
\affil{Division of Physics, Mathematics and Astronomy, California Institute of Technology,
Pasadena, CA 91125}
\and
\author{Weidong Li, Alexei V. Filippenko}
\affil{Department of Astronomy, 601 Campbell Hall, University of California, Berkeley, CA 94720-3411}


\altaffiltext{1}{Hubble Fellow.}



\begin{abstract}

We had previously reported on a survey for radio
transients, conducted by comparing the FIRST and NVSS radio
catalogs, and resulting in the discovery of nine possible
sources. These were used to
set an upper limit on the number of orphan gamma-ray burst (GRB) radio
afterglows, and thus a lower limit on the typical GRB beaming
factor ($f_b^{-1} \equiv (\theta_{jet}^2/2)^{-1}$). 
Here we report radio and optical follow-up observations 
of all of these possible transients, achieving the first full 
characterization of the transient radio sky. We find that seven
source are unlikely to be real radio transients. Of the
remaining two sources, one is most probably an optically obscured
radio supernova (SN) in the nearby galaxy NGC 4216, the first such
event to be discovered by a wide-field radio survey. The second source
appears not to be associated with a bright host galaxy (to a limit of
$R < 24.5$ mag). Too radio luminous to be associated with a GRB,
we speculate that this may be a flare from a peculiar, variable, radio-loud 
active galactic nucleus, or a burst from an unusual Galactic compact object, 
but its exact nature remains a mystery and merits further study. 
We place an upper limit of 65  
radio transients above 6 mJy over the entire sky at the $95\%$ confidence level.  
The implications are as follows. First,
following the analysis in our previous paper, 
we derive a limit on the typical beaming of GRBs; 
we find $f_b^{-1} \simgt 60$,
$\sim5$ times higher than our earlier results. 
Modifying our model parameters
and analysis scheme leads to values
which are likely to be within an order of magnitude from this number. 
Second, our results impose an upper limit
on the rate of events that eject $\gtrsim 10^{51}$ erg in 
unconfined relativistic ejecta, such as conical jets, 
whether or not accompanied by detectable emission in wavebands
other than the radio. Our estimated rate, $\dot{n}\leq1000$ yr$^{-1}$ 
Gpc$^{-1}$, is about two orders of magnitude 
smaller than the rate of core-collapse SNe (and type Ib/c events
in particular), indicating that only a minority of such events
eject significant amounts of relativistic material, which are required
by fireball models of long-soft GRBs.
Finally, we consider the prospects of future radio surveys. 
We show that future wider and/or deeper radio variability
surveys are expected to detect numerous orphan radio GRB
afterglows. Furthermore, the fact that our survey probably
detected an optically obscured radio SN illustrates the great
potential of sensitive surveys with new radio instruments
to revolutionize the study of nearby SNe.

\end{abstract}




\section{Introduction}

Exploring the time domain with wide-field surveys, covering
significant parts of the sky, is one of the promising new 
frontiers in observational astronomy. This area has been little
explored so far, mostly due to the technical difficulty in conducting
multi-epoch deep surveys which cover wide areas of sky.  Opening this
new window of discovery is one of the main objectives of new, large optical
surveys, like the Palomar QUEST Survey (Djorgovski et
al. 2004) and the Supernova Legacy Survey (SNLS, Sullivan et al. 2004),
and is a major science driver for more ambitious forthcoming
initiatives such as Pan-STARRs (Kaiser 2004), the Large Synoptic
Survey Telescope (LSST, Claver et al. 2004), and the SuperNova
Acceleration Probe (SNAP, e.g., Linder et al. 2004). Initiatives 
in the radio band include the Allen Telescope 
Array (ATA) and the Square Kilometer Array
(SKA, Carilli \& Rawlings 2004). Some initial results from the
above-mentioned optical surveys are already emerging (e.g., QUEST,
Mahabal et al. 2004; and SNLS, Sullivan et al. 2004), and some relevant
studies are also being conducted using the Sloan Digital Sky Survey (SDSS;
e.g., Lee et al. 2003) even though, by design, this survey invests
few resources in exploring the time domain.

Large parts of the high-energy (gamma-ray and X-ray) sky
have been almost continuously monitored for variability in the last
decades, by dedicated space missions such as the Rossi X-ray Timing
Explorer, the BATSE instrument on board the Compton Gamma-Ray
Observatory, and the interplanetary network (Hurley et al. 2002), but
this wide-field monitoring is conducted by low-resolution instruments
(typically worse than $1'$), making the discovered transients difficult to
localize.  While higher-resolution X-ray imaging is possible with
instruments on board {\it Swift}, {\it Chandra}, and {\it XMM-Newton}, the limited
field of view and long exposures required for imaging faint sources
makes sensitive, high-resolution, wide-field surveys in these bands
impractical (although see Read et al. 2004).

In the radio band, wide-field surveys covering most of the sky have
been carried out. While not originally designed for variability
studies, these surveys do offer the opportunity to explore the time
domain over a significant part of the sky, with reasonable sensitivity
(a few mJy) and at decent resolution (significantly better than $1'$). 
Indeed, in Levinson et al. (2002, hereafter Paper I) we carried out such an
analysis by comparing two wide-field surveys conducted with the Very
Large Array (VLA) -- the ``Faint Images of the Radio Sky at Twenty centimeters" 
survey (FIRST, White et al. 1997) and the NRAO VLA Sky Survey (NVSS,
Condon et al. 1998) -- in search of transient sources. 

That study was motivated by the
search for a specific phenomenon, the so-called ``orphan'' afterglows
of cosmological gamma-ray bursts (GRBs; e.g., Rhoads 1997). If the radiation
associated with a GRB is beamed (i.e., emitted into a small solid angle)
it follows that we do not observe many such events whose radiation is beamed 
away from us. At late times, the emitting material decelerates and the 
lower-energy afterglow emission from such GRBs (e.g., in the radio), becomes 
isotropic, and therefore observable (see Paper I for more details). 
We thus expect ``orphan'' GRB radio afterglows
to appear as transient radio sources which are not related to any observed GRB. 
Our survey revealed nine possible radio transients, and in
Paper I we used this to investigate the typical beaming of GRBs. 

Here, we report the results of a follow-up effort that leads to the full
characterization of this sample, and thus of the transient radio sky. 
A plan of the paper follows. We describe radio and optical follow-up observations 
of the candidate radio transients from Paper I in $\S~2$. In section $\S~3$ we report on the 
properties of the two real transient sources we discovered, a probable radio dupernova 
(SN) in a nearby galaxy and a source with no optical counterpart which we show is unlikely
to be associated with a GRB. In $\S~4$ we discuss the main implications of our work,
including an improved limit on the typical beaming of GRBs ($\S~4.1$), and a limit
on the total rate of nearby relativistic explosions, which implies that most core-collapse 
SNe do not eject unconfined relativistic outflows ($\S~4.2$).  
We also take a broader approach, and discuss the implications
of our findings in the context of current and future wide-field
variability surveys in ($\S~4.3$), and summarize our results in ($\S~4.4$).

\section{Observations}

\subsection{VLA Follow-Up of the Radio Transient Candidates}

Following the discovery of nine candidate radio transients in the
survey described in Paper I, we launched a follow-up program with the
Very Large Array (VLA{\footnotemark\footnotetext{The VLA is
    operated by the National Radio Astronomy Observatory, a facility
    of the National Science Foundation operated under cooperative
    agreement by Associated Universities, Inc.}})  at both 1.43 GHz
and 8.46 GHz.  The initial observations toward all nine sources were
taken in 2002 May and 2002 November with the VLA in the BnA, B, and C
array configurations -- matching the resolution of the FIRST
survey. The integration times were chosen so that any bona fide orphan
afterglow candidates could be detected on the basis of their power-law
decline.  Dual-frequency observations were used to obtain some spectral
discrimination for identifying variable, flat-spectrum active galactic
nuclei (AGNs). The data were reduced with the Astronomical Image Processing
Software (AIPS) in the standard manner. In Table 1 we present a
summary of the available radio observations for each of these
candidates. For one source (\#4 in Table 1.; VLA\,121550.2+130654) we made additional
flux-density observations and re-analyzed available archival
data. These data are summarized in Table \ref{tab:data}; see \S~2.2
for more details.

Of the nine objects identified in Paper I, we find that five are
unlikely to be variable at all. Of these, two (\#1 and \#3 in Table 1)
are constant sources
whose flux measurements were compromised by side lobes from nearby
bright sources (flux $\ge1$ Jy). Two others (\#2 and \#7 in Table 1) 
are constant sources that
were not properly deblended from neighboring objects in the
lower-resolution NVSS survey. The final false candidate 
(\#6 in Table 1) appears to be
an image artifact of an unexplained nature in the FIRST catalog.

Of the remaining four candidates, two are variable
sources whose nature is of little interest to our current
survey. VLA122532.6+122501 (\#5 in Table 1) is a radio-variable, flat-spectrum source
($\alpha=-0.16$) projected on the nucleus of a nearby galaxy (VPC 418;
$R \approx 16$ mag; Young \& Currie 1998), and thus most likely an 
AGN. VLA165203.1+265140 (\#8 in Table 1) is coincident with the
known pulsar PSR J1652+2651.
 
The last two sources are interesting, and remain viable radio
transient candidates. VLA121550.2+130654 (\#4 in Table 1)
is a non-nuclear
steep-spectrum source in the nearby galaxy NGC 4216.  This source
continued to brighten, then peaked during our observations, and Figure 1 shows the
light curve compiled from all available data. Our modelling (e.g., Soderberg, Nakar,
\& Kulkarni 2005a) shows that the location and
light curve of this radio transient are consistent with it being a
sub-relativistic Type II supernova (SN II) in NGC 4216, but they are
also consistent with the relativistic ejecta resulting from an
off-axis GRB. 
Very Long Baseline Array (VLBA) observations
designed to measure the spatial size of the remnant and thus deduce the
speed of the ejecta, discriminating 
between these two classes of stellar explosions, are described in the
next section.

VLA172059.9+385227 (\#9 in Table 1)
was clearly detected (with a flux of $9.4 \pm 0.2$
mJy) in a FIRST survey image obtained in August 1994 (detection limit
$\sim1$ mJy), but is absent from an NVSS image obtained in April 1995
with comparable sensitivity, as well as in our subsequent imaging during
2002. Thus, this source appears to be a truly transient radio
source. The absence of a bright host galaxy disfavors a radio SN
identification for this source, but it remains a viable GRB radio
afterglow candidate. Below we explore this further using deep imaging of the
location of VLA172059.9+385227.

\subsection{Investigation of VLA121550.2+130654}

\subsubsection{VLBA Observations of VLA121550.2+130654}

To measure the size of the radio source, we obtained an 8-hr
VLBA observation of VLA121550.2+130654 on 2004 May 2 UT.
The observation was taken in standard continuum mode with a bandwidth of
$4\times 8$ MHz centered on observing frequencies of 1.4 and 8.5 GHz.
Fringe calibrations were applied using 3C286 and phase referencing was
conducted using J1207+1211 at an angular distance of $2.3^{\circ}$ from
the radio transient.

We detect the source at a position of $\alpha=12^{\rm h} 15^{\rm m}
50^{\rm s}.235$, $\delta=+13^{\circ} 06' 54''.03$ (ICRS J2000.0) with a
positional uncertainty of 10 mas in each coordinate
(Figure~2).  We note that these errors are dominated by
the positional uncertainty of J1207+1211.  Using the VLBA utilities
within AIPS, we find a flux density for the transient of $F_{\nu,
  1.4\rm~GHz}=9.63 \pm 0.40~\rm mJy$ and $F_{\nu, 8.5\rm
  ~GHz}=2.45\pm 0.55$ mJy. Both 1.4 GHz and 8.5 GHz detections are essentially 
unresolved within the beams, $9.53 {\rm ~mas} \times 4.91$ mas at 1.4 GHz 
and $1.82 {\rm ~mas} \times 0.91$ mas at 8.5 GHz.
Analysis by VLBA custom software including circular and elliptical Gaussian fits to the
data result in $3\sigma$ upper limits on the source diameter of $2.9(4.5) {\rm ~mas}$ 
at 1.4 GHz, and $3.4(4.0) {\rm ~mas}$ at 8.5 GHz for
circular (elliptical) Gaussian models, respectively (M. Bietenholz, 2005,
private communication). 
We note that there is no emission from the host galaxy at this
resolution.  

\subsubsection{Optical Monitoring of NGC 4216, Host Galaxy of VLA121550.2+130654}

In order to detect, or set limits on, any optical emission coincident
with the emergence of this radio source, we have examined optical
images of NGC 4216, obtained between the years 1991 and 2004. We
retrieved images obtained by the 48-inch (as part of the second Palomar
sky survey, POSS-II) and 60-inch telescopes at Palomar Observatory
from the NASA Extragalactic Database (NED)\footnotemark,
\footnotetext{The NASA/IPAC Extragalactic Database (NED) which is
  operated by the Jet Propulsion Laboratory, California Institute of
  Technology, under contract with the National Aeronautics and Space
  Administration.}  and images obtained at the JKT telescope at La
Palma from the ING archive\footnotemark \footnotetext{\url
  http://archive.ast.cam.ac.uk/ingarch/ .}. We have also re-examined
numerous images obtained by the Lick Observatory Supernova Search 
utilizing the 30-inch Katzman Automatic Imaging Telescope (KAIT;
Li et al. 2000; Filippenko et al. 2001; Filippenko 2005)
at Lick Observatory, between April 1997 and May
2004. All these images reach comparable depth ($\sim19$ mag) and do
not show a compact source at the radio position of
VLA121550.2+130654. Some of the best KAIT data were also intercompared
using the CPM image subtraction method (e.g., Gal-Yam et al. 2004),
and we detect no variable optical source at the radio location down to
the KAIT detection limit, typically $R=19.5$ mag.
 
\subsection{Optical Follow-Up of VLA172059.9+385227}

As reported in Paper I, inspection of Palomar Digital Sky Survey plate
data covering the location of VLA172059.9+385227 did not reveal any
candidate host galaxies of this event. We have therefore obtained 
deeper optical imaging using the Palomar Observatory 200-inch telescope 
(P200; $R$ band) and the Keck-I 10-m telescope ($I$ and $g$ bands). 

Photometric calibration of this field, using Landolt (1983) standard stars, 
was obtained under photometric conditions with the Wise 
Observatory 1-m telescope. Six secondary calibrators in the
vicinity of VLA172059.9+385227 were measured, with typical magnitudes
of 19 (in the $R$ band). However, as these stars were saturated in our
deeper Keck and P200 images, additional, intermediate-depth images
were collected with the robotic 60-inch telescope at Palomar
(Cenko et al. 2005, in prep.), and
used to place unsaturated objects in our deep images on the 
zero points defined by the Wise calibration. 
An observing log is given in Table 3. 

\section{Results}

\subsection{VLA121550.2+130654: a Likely Radio Supernova in NGC 4216}

Figure 1 and Table 2 show the radio (1.4 GHz) light curve of this
source. We find that its overall characteristics are consistent with
those of a Type II SN (see Weiler et al. 2002 for
a review).  In particular, the temporal evolution and the steep
spectral index are consistent with those measured for the radio-luminous 
Type II SN 1979C (Weiler et al. 1991). To illustrate this,
plotted in red (dashed curve) in Figure 1 is a continuous model curve which has been
shown by Weiler et al. (1991) to describe the 1.4 GHz observations of
SN 1979C very well. The plotted curve is shifted in time and scaled in
flux to best fit the data, but the shape of the curve is kept constant
(i.e., we applied no ``stretch'' correction). As can be seen, this
curve describes our data quite well. While some Type Ic SNe are also
radio bright (e.g., SN 1998bw, Kulkarni et al.  1998; SN 2003L,
Soderberg et al. 2004), their light curves evolve quickly and are
inconsistent with our data, as demonstrated by the cyan (solid) curve in Figure
1, representing the radio light curve of SN 1998bw from Kulkarni et
al. (1998) scaled in flux and time as above. This light curve needs
to be ``stretched'' by a factor of $\sim50$ in order to match the
temporal evolution of VLA121550.2+130654.

NGC 4216 is a member of the Virgo cluster of galaxies. Assuming a
distance of 15.9 Mpc to the Virgo cluster (Graham et al. 1999), the
peak flux of this event was $\sim2.7\times 10^{27}$ erg s$^{-1}$ Hz$^{-1}$, 
also typical of known radio SNe (Weiler et al. 2002).
 
At the bottom of Figure 1 we mark the periods of time in which a
bright optical SN in NGC 4216 would have been visible in the archival
images we have collected ($\S~2.2.2$).  The black arrows mark the
actual dates of observations. The horizontal line is our estimate for
the range of dates in which an unobscured SN with peak optical
luminosity and light-curve shape similar to those of SN 2002ap (e.g., Gal-Yam,
Ofek, \& Shemmer 2002; Foley et al. 2003), and which would have been
visible in these archival data, would reach peak brightness.  In other
words, SNe whose peak brightness occurred typically between 250 days before
and 10 days after each observation would have been detected in these
images.  Optically luminous SNe with broader light curves (e.g., radio-bright 
events like SN 1979C, SN 1998S, and SN 1998bw) would have been visible
for even longer periods of time. The fact that we do not detect an
optical counterpart to VLA121550.2+130654 in any of the images we
inspected, combined with the effectively continuous monitoring of this
galaxy in the last decade, suggests that this event was probably
heavily obscured by dust.

\subsubsection{Angular Size of the Radio Ejecta}

If VLA121550.2+130654 is a Type II supernova, then we can estimate its expected
angular diameter by comparing it with other well-studied supernovae.
We chose to compare it with SN 1979C since both objects are of
comparable brightness, and the host galaxy of SN 1979C (M100) and the
host galaxy of VLA121550.2+130654 (NGC 4216) both lie in the Virgo cluster.  
VLBI measurements by Bartel \& Bietenholz (2003) taken from
3.7 to 22 years after the explosion of SN 1979C show an almost free expansion over
this time. Adopting their best-fit parameters, 
the expected angular diameter for an isotropic
expansion is $\theta$=2.1 ($t_{\rm years}$/7.0) mas. 

In contrast, if VLA121550.2+130654 is a GRB we estimate its angular diameter
(Frail, Waxman, \& Kulkarni 2000) at the distance of NGC 4216 to be
$\theta=17\,(E_{51}/n_\circ)^{1/5}(t_{\rm years}/7.0)$ mas, where $E_{51}$
is the kinetic energy of the shock and $n_\circ$ is the density of the
circumburst medium. This estimate assumes an isotropic outflow
expanding non-relativistically (i.e. Sedov-Taylor dynamics). In
reality, during the early phase ($t<0.5$ yrs) the GRB outflow expands
relativistically and the geometry is probably jet-like. More detailed
calculations for the size evolution of GRB jets for different viewing angles
is given in Granot \& Loeb (2003). Note that both SN and
GRB explosions release about 10$^{51}$ erg of kinetic energy but the
first drives a slow shock ($v_s \approx 5000$ km s$^{-1}$), while the other
drives a shock that is initially relativistic ($v_s\approx c$).

In the case of VLA121550.2+130654, our upper limits on the angular size are
consistent with the size expected from scaling the observations 
of SN 1979C, but are in strong conflict with the predictions from
GRB models. We therefore conclude that
VLA121550.2+130654 was a radio-selected Type II supernova.

\begin{figure*}
\includegraphics[width=15cm]{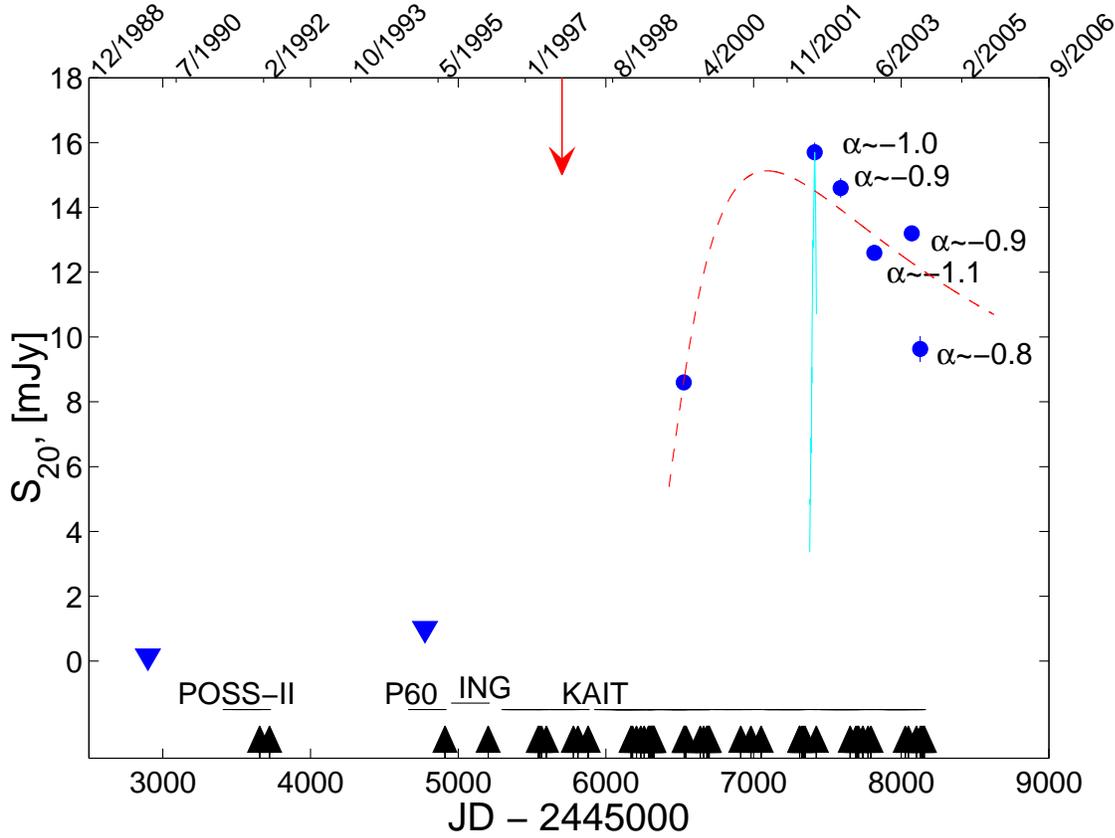}
\caption{Radio (1.4 GHz) light curve of VLA121550.2+130654. Inverted triangles
  mark upper limits from archival VLA observations. Superposed are the
  radio light curves of two radio-bright events: SN 1979C (Type II;
  Weiler et al. 1991; dashed red curve) and SN 1998bw (Type Ic; Kulkarni et
  al. 1998; solid cyan curve). These were scaled in flux and shifted to match the
  approximate time of peak radio luminosity (see text). The light
  curve of VLA121550.2+130654 is quite similar to that of SN 1979C,
  and markedly different from that of SN 1998bw, suggesting a Type II
  identification for this event. The red arrow at the top of the
  figure marks the date of peak optical brightness of SN 1979C,
  relative to its peak 1.4 GHz radio flux. The bottom of the
  figure describes a search for a bright optical SN in NGC 4216 in
  archival images we have collected, obtained by the 30-inch KAIT, the
  Palomar 60-inch and 48-inch (POSS-II) telescopes, and the 1-m JKT telescope.  The
  black arrows mark the dates of observations while the horizontal
  lines represent our estimate for the period of time in which an
  unobscured SN would have been visible in these archival data (see
  text).  An optical counterpart to VLA121550.2+130654 is not detected
  in any of the images we inspected, suggesting that this event was
  heavily obscured by dust.}
\end{figure*}

\begin{figure*}
\includegraphics[width=17cm]{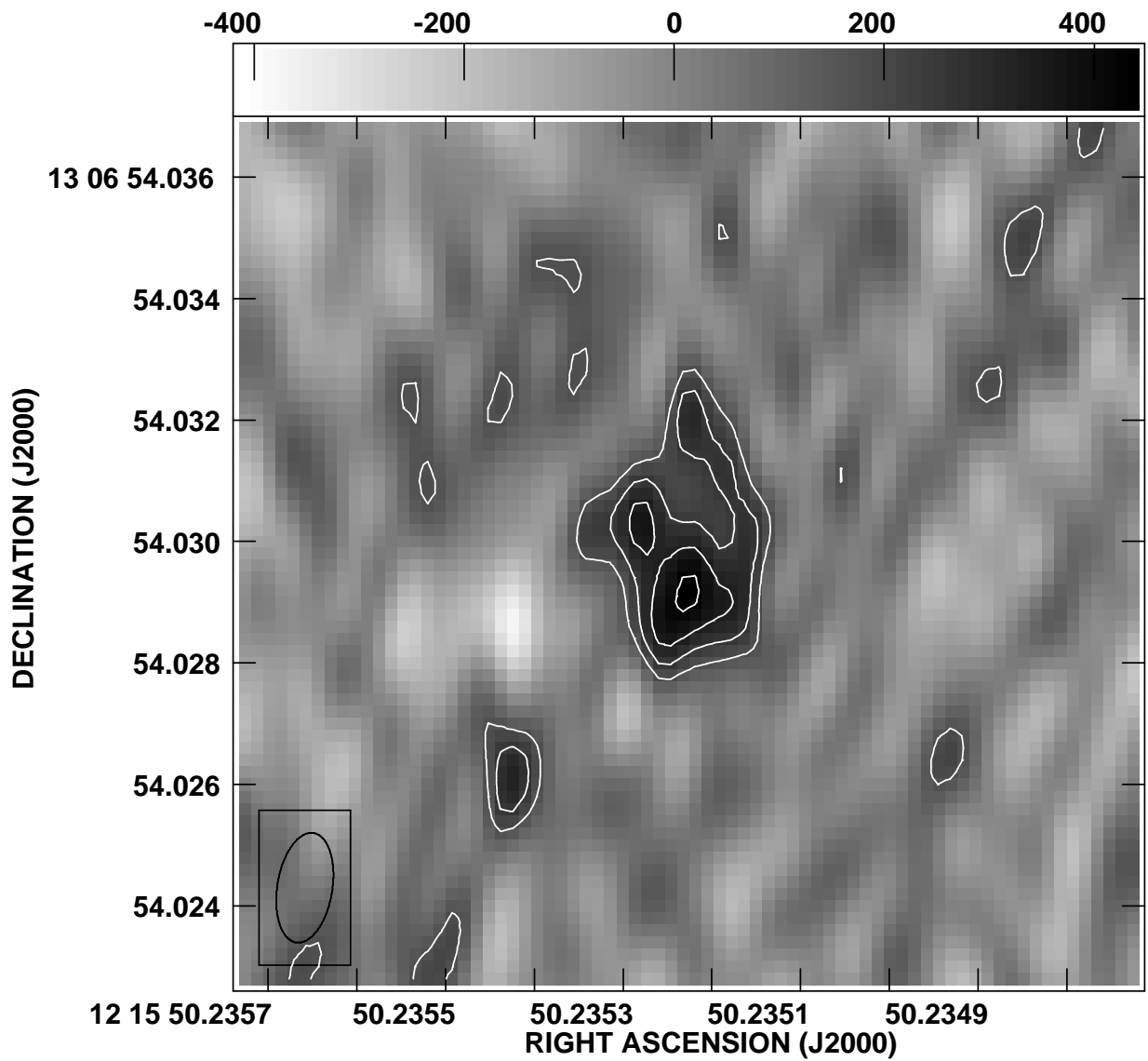}
\caption{VLBA contour map of VLA121550.2+130654 at $t\approx 7$ years. At
  8.5 GHz, the radio transient may be marginally resolved with a slightly
  asymmetric structure. Gaussian fitting yields size estimates that are 
  comparable to the size of the synthesized beam, shown in the lower left-hand
  corner. We place a firm $3\sigma$ upper limit of $4.0 {\rm ~mas}$ on the
  source size (see text). The greyscale intensity map spans from $-408.3$ to 436.4
  $\mu$Jy beam$^{-1}$. Contours represent flux in linear increments
  from 2$\sigma$ to 10$\sigma$ (84 to 840 $\mu$Jy).}
\end{figure*} 

\subsection{VLA172059.9+385227: A Radio Transient Source Without 
a Bright Optical Counterpart}

Figure 3 shows the location of VLA172059.9+385227 on our deepest
$R$-band image.  The image astrometry was solved with respect to the
USNO-A2.0 catalog (Monet et al. 2003) using 10 unsaturated stars, and has 
a root-mean-square (RMS) accuracy
of $\sim0.4''$. Cross-correlating the positions of 3300 sources
within $10^{\circ}$ from VLA172059.9+385227 which appear in both the
FIRST and USNOA-2.0 catalogs, we derive the total uncertainty in
placing FIRST sources (with similar flux to that of
VLA172059.9+385227) on the USNOA-2.0 reference frame. This error
includes both statistical uncertainties in the reported positions of
cataloged sources, as well as any systematic deviations between the
FIRST and USNOA-2.0 reference frames. We find the total uncertainty
for this source to be $0.72''$ at the $1\sigma$ confidence
level.  Reassuringly, $\chi^2$ analysis of the residuals shows that
the systematic difference between the FIRST and USNO reference systems
at this location must be small ($<0.1''$). Accordingly, the radius of
the circle marked on Figure 3 is $0.72''$, demonstrating that there is
no galaxy detected in the vicinity of the radio source. The nearest
galaxy (marked A in Fig. 3; total $R$-band magnitude $\sim24.5$) is
$\sim3''$ away. We therefore determine that any point source
or compact galaxy at the location of VLA172059.9+385227 must be
fainter than $R=24.5$ mag. Similar limits are obtained from our
$g$-band and $I$-band observations.  We cannot firmly rule out an
association between VLA172059.9+385227 and nearby galaxies A, B, or C,
since our ground-based imaging lacks the depth and resolution required
to properly model the light distribution of these faint sources, and
so determine how likely is the association of the radio source with
these galaxies (as done, e.g., by Gal-Yam et al. 2003). 
However, the density of similar sources in our deep
images suggests that chance coincidence is quite possible.

As shown in Paper I, the typical distance to an orphan radio afterglows
detected in our survey, assuming our fiducial parameters, should be $\le140$ Mpc
($z\approx 0.033$ for $H_0 = 70$ km s$^{-1}$ Mpc$^{-1}$), and even under the
most favorable assumptions these events should always be below $z\approx0.2$. 
At that distance any possible
host galaxy (either galaxy A or another undetected galaxy
closer to the radio location) would have an absolute magnitude
$M_R>-11$ for $z=0.033$, or $M_R>-15.5$ for $z=0.2$, i.e., be a very
low luminosity dwarf (fainter, possibly much fainter, than the Small Magellanic
Cloud). This leads us to conclude that this source is unlikely to have
been an orphan radio GRB afterglow. 

If not an orphan GRB afterglow, what is the nature of this source? We now
briefly consider several alternatives. In principle, this could have been an
on-axis GRB afterglow, as these are seen to great cosmological
distances, and often reside in host galaxies which have very faint
apparent optical magnitudes (e.g., Vreeswijk et al. 2001; Jaunsen et al. 
2003; Berger et al. 2002; Berger et al. 2005, in prep.).
However, the radio brightness of this transient (9.4 mJy) is unprecedented
for on-axis GRBs; it is far brighter than every 
afterglow of cosmological GRBs observed to date (e.g., Frail et al. 2003).
In addition, in paper I it is shown that the population of observed radio afterglows
is always dominated by those that have become almost spherical. Thus, if this source
is a distant beamed afterglow, we would have expected to see many other
less-beamed afterglows, which we don't. 
We thus conclude that this source is unlikely to be associated with a GRB,
either on-axis or off-axis.

This source could have been a radio flare from a
peculiar AGN, perhaps an extreme and/or high-$z$ analog of SDSS
J124602.54+011318.8 (Gal-Yam et al. 2002). If that is the case, then
during the flare caught by the FIRST observations, the radio loudness
of this AGN was extremely high ($\Re > 10000)$\footnotemark.
\footnotetext{Following Stocke et al. (1992), we define the radio
  loudness $\Re$ as the ratio of the radio flux at 5 GHz to the optical
  $B$-band flux. We translate our $R$-band upper limit and 1.4 GHz
  radio flux to the Stocke et al. bands assuming a typical flat AGN
  spectral slope ($\alpha \approx 0.5$) and a source redshift
  $z<3$. Assuming higher redshifts for the source would increase $\Re$,
  by up to an order of magnitude at $z=5$. Assuming steeper spectral
  slopes would decrease $\Re$, by up to an order of magnitude for
  $\alpha=1.6$, which is quite atypical for AGNs.}  

Alternatively, we may have observed a radio flare from a Galactic object,
perhaps similar to those recently reported by Hyman et al. (2005) and
Bower et al. (2005). Possible sources for such transient flares are 
discussed by these authors, as well as by Kulkarni \& Phinney (2005)
and Turolla, Possenti, \& Treves (2005). However, the high Galactic latitude
of this source ($\sim33^{\circ}$) appears to disfavor this option.

Finally, since ours is the first wide-field survey for radio 
transients, we may have discovered a new type of object, yet to be
characterized. To conclude, the
exact identification of VLA172059.9+385227, the last event in our
survey which we cannot yet securely associate with a known astrophysical source,
remains somewhat of a mystery, and definitely merits further
study. However, it appears that this source is not at low redshift, 
and is therefore unlikely to be a radio orphan GRB afterglow.
 
\begin{figure*}
\includegraphics[width=17cm]{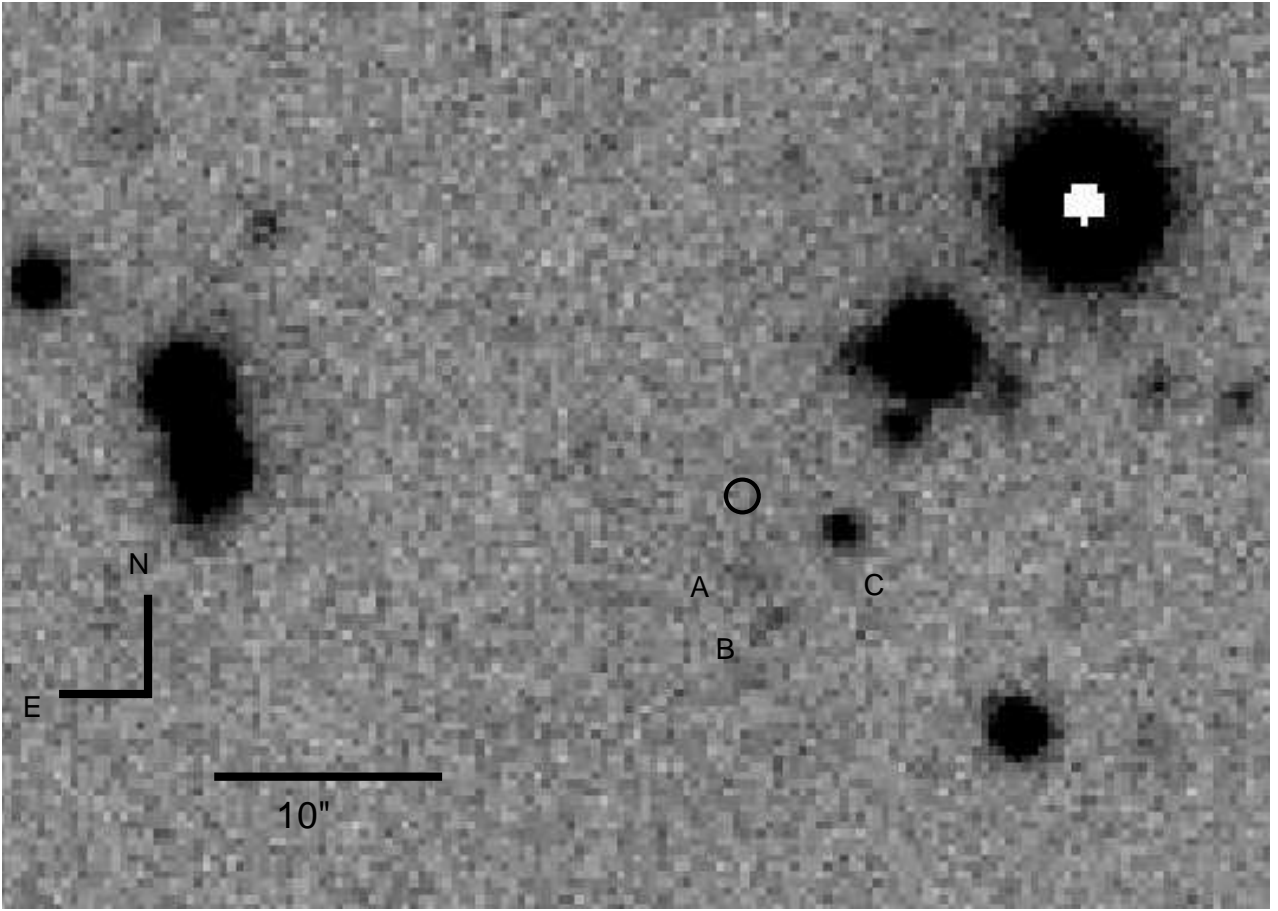}
\caption{P200 $R$-band image of the location of
  VLA172059.9+385227. Nearby galaxies A ($R=24.5\pm0.4$ mag) B
  ($R=24.6\pm0.4$ mag) and C ($R=23.5\pm0.4$ mag) are marked. 
  The bright source due North-West is a star. 
  The radio location is marked by the
  black circle, with $0.72''$ radius representing the $1\sigma$
  positional uncertainty. The distance to the nearest detected galaxy
  (A) is $\sim3''$. The faintness of all possible host galaxies argues
  against a low-$z$ origin for VLA172059.9+385227, and therefore
  against it being an orphan radio GRB afterglow.}
\end{figure*} 

\section{Discussion and Conclusions}

\subsection{Limits on the Typical Beaming of GRBs}

In Paper I it has been shown that for a GRB population with a given
isotropic equivalent burst energy, $E_{\rm iso}$, the number of orphan
radio afterglows anticipated to be detected in a flux-limited survey
is smaller for larger beaming factor $f_b^{-1}$, contrary to naive
expectations (here $f_b \equiv \theta^2/2$, with $\theta$ being the opening
angle of the GRB ejecta). Obviously, if the beaming factor is larger,
more GRBs occur in nature, as the rate measured by earth-orbiting spacecraft 
represents a smaller fraction of the total population, most of which 
is beamed away from us. However, since the energy we measure is  
$E_{\rm iso}$, which is related to the true
energy of the bursts by $E_b=f_b E_{\rm iso}$, a larger beaming factor
implies that the typical burst is less energetic. This will cause the
number of observed afterglows to be smaller for two reasons. 
First, a smaller true energy $E_b$ implies a
smaller luminosity distance below which the radio flux emitted by a
source that has undergone a transition from relativistic to
subrelativistic expansion exceeds some detection limit. Second, the
time a source spends above the detection limit is shorter for a
smaller $E_b$. In Paper I we have shown that these two effects
combined overcome the expected increase in source counts due to 
the larger true GRB rate inferred from the observed rate for
larger beaming factors.

For a flux threshold of $6$ mJy, as in the present
analysis, the maximum redshift below which sources are above the
detection limit is $z \approx 0.2$ for $h=0.75$ (Paper I;
$H_0 = 100h$ km s$^{-1}$ Mpc$^{-1}$), and so
cosmological effects can be neglected.  
In this limit the number of
radio orphans expected in a survey is proportional to $f_b^{5/6}$
(Paper I).  Thus, the upper limit derived on the number of radio
afterglow sources implies a lower limit on the beaming factor,
$f_b^{-1}$.  In
Paper I we obtained $f_b^{-1}>13$ at the 95\% confidence limit (CL) 
using a complete subsample
out of the 9 candidates that were identified there, a local GRB rate of
$\dot{n}=0.5$ Gpc$^{-3}$ yr$^{-1}$, and our canonical choice of the
remaining parameters.  A recent analysis by Guetta, Piran, \& Waxman
(2005) yields a local GRB rate of $\dot{n}=0.67 (h/0.75)^3$ Gpc$^{-3}$ yr$^{-1}$.  

The rejection of all candidates by the follow-up observations described 
in this paper implies an upper limit of 65 all-sky radio afterglows
above 6 mJy at 95\% Poisson CL (see Paper I for further details).
With this new upper limit, and using eq. (9) of Paper I, the
modified lower limit on the beaming factor is
\begin{equation}
f_b^{-1}\ge 62\left(\frac{\dot{n}}{0.67~{\rm Gpc}^{-3}~{\rm yr}^{-1}}\right)^{6/5}
\left(\frac{\epsilon_B}{0.03}\right)^{27/20}\left(\frac{\epsilon_e}{0.3}\right)^{9/10}
\left(\frac{n}{0.1~\rm cm^{-3}}\right)^{19/20}
\left(\frac{\tilde{E}}{5\times 10^{53}~\rm erg}\right)^{11/5},
\label{f>}
\end{equation}
where $\epsilon_B$ and $\epsilon_e$ are (respectively) the magnetic field 
and relativistic electron equipartition fractions, 
$n$ is the density of the local ambient medium in which the blast 
wave propagates, and $\tilde{E}\equiv<E_{\rm iso}^{11/5}>^{5/11}$.  

Model parameters are normalized in eq.~(\ref{f>}) to values 
derived from afterglow observations.  Values of close to
equipartition, $\epsilon_e\ge0.1$, are typically inferred from most
optical afterglows, and from the clustering of explosion energies
(Frail et al. 2001) and X-ray afterglow luminosity (Freedman \& Waxman
2001; Berger, Kulkarni, \& Frail 2003); $\tilde{E}$ is normalized to
the value $<E_{\gamma,\rm iso}^{11/5}>^{5/11}=5\times10^{53}$~erg
obtained for GRBs with known redshifts (Bloom et al. 2003), taking
into account that while the GRB kinetic energy is likely several times
larger than the $\gamma$-ray energy, the population of GRBs with
known redshifts is probably brighter, on average, than the whole
cosmological GRB population.

The values of $n$ and $\epsilon_B$ are not as well determined by
observations as the values of $\tilde{E}$ and $\epsilon_e$. The reason
is that they depend strongly on an observational parameter which is
less well determined by afterglow observations: the self-absorption
frequency $\nu_a$. While $\tilde{E}\approx 1/\epsilon_e \approx
\nu_a^{(5/6)}$, we have $n\approx \nu_a^{(25/6)}$, and $\epsilon_B\approx
\nu_a^{(-5/2)}$. The value of $\nu_a$ is determined at best, in only a few cases,
to within a factor of 2--3. 
Therefore, in most cases the uncertainty in determining $n$ (and
$\epsilon_B$) is at least an order of magnitude.  In cases where
$\epsilon_B$ can be reliably constrained by multi-waveband spectra,
values not far below equipartition are inferred (e.g., Frail, Waxman,
\& Kulkarni 2000).  Our analysis depends on 
$\epsilon_B^{27/20} \times n^{19/20} 
\approx \nu_a^{7/12}$, so we are not so sensitive to
the uncertainty in $n$ and $\epsilon_B$ separately.  As explained in
Paper I, afterglow observations typically imply $\epsilon_B \times
n\ge 10^{-3}~{\rm cm^{-3}}$.

It should be
emphasized, however, that the lower limit of eq.~(\ref{f>}) is
uncertain due to uncertainties in model parameters. Afterglow
models are highly idealized and the values of model parameters are
therefore accurate only to within a factor of a few. 
Nevertheless, this analysis provides a direct evidence for beaming 
in GRBs which is independent of that provided by afterglow light curves.

The lower limit directly imposed on $<f_b^{-1}>$ 
by our analysis is consistent with the value $<f_b^{-1}>=75\pm25$ derived by 
Guetta et al. (2005). These authors compare the Frail et al. (2001) distribution 
of jet opening
angles inferred from breaks in the afterglow light curves with model predictions
applied to the BATSE GRB catalog, and their results are thus completely independent.
This is encouraging and has two major implications. First, if this consistency 
is to be maintained, future deeper or wider radio surveys should 
detect many afterglows ($\S~4.3$). Second, 
while the values of $\epsilon_B$ and $\epsilon_e$ may be assumed
universal, as they are determined by shock microphysics, $n$ may vary
significantly among bursts. Higher values for the typical ambient
density ($n\approx10~{\rm cm^{-3}}$), as advocated by Bloom, Frail, \&
Kulkarni (2003), require low values of $\epsilon_B$ in order to
reproduce afterglow observations (Paper I).  A combination of large
$n$ and large $\epsilon_B$ would drive $<f_b^{-1}>$ to large values
(eq.~\ref{f>}), resulting in a strong inconsistency with the analysis
of Guetta et al. (2005).  Recast in another way, requiring consistency
between our results and the independent analysis of Guetta et
al. (2005) could be taken as an indication that values of $\epsilon_B
\times n/(0.01~{\rm cm}^{-3}) >> 1$ are ruled out.

Note that this analysis assumes no correlation between $E_{\rm iso}$ and 
the beaming fraction $f_b$, which is the most general assumption that can 
be made. Our results will be modified by factors of order a few if a 
correlation is assumed, depending on its exact form.
For example, if we assume the
correlation derived by Frail et al. (2001), namely a constant $E_{\rm iso} \times f_b$,
then we should replace $<E_{iso}^{11/6}>$ by $<E_{iso}>^{11/6}$ in eq. (9) of Paper I, 
resulting here in a modified lower limit $f_b^{-1}>20$. In fact, 
this is the value that should be compared with the analysis of Guetta et 
al. (2005), which explicitly assumes the Frail et al. (2001) correlation.
This agreement ($f_b^{-1}>20$ compared to $f_b^{-1}>=75\pm25$) shows
that our above conclusions are supported also in this case. 

\subsection{An Upper Limit on the Total Rate of Relativistic Explosions}

Little is known about the fraction of relativistic cosmic explosions that 
produce bright $\gamma$-ray radiation. Observationally, the existence of
X-ray flashes (XRFs; Heise et al. 2001) and their association with Type Ib/c
SNe (Soderberg et al. 2005b) suggests that the peak energy of
such explosions can be at soft X-rays, or even in the very 
far ultraviolet, in which case
they will be very difficult to detect. 
For example, a ``dirty'' fireball, with a Lorentz factor low enough so it is
optically thick to $\gamma$-rays, will escape real time detection by
orbiting $\gamma$-ray observatories (Rhoads 2003). 
However, all such explosions, regardless of the explosion geometry
and the initial Lorentz factor (as long as it is $ \simgt 2$), would produce
similar late radio afterglows.  

Adopting the parameters values of Paper I and our upper limit on the number
of relativistic explosions ($<65$ over the entire sky) we can write eq. (9) 
from Paper I as

\begin{equation}
\dot{n}\simlt 1000 \times E_{0,51}^{-11/6} {\rm Gpc}^{-3} {\rm yr}^{-1},
\label{skyrate}
\end{equation}

\noindent
Where $E_{0,51}$ is the total energy in relativistic ejecta in units
of $10^{51}$erg, and the propagated confidence level from the upper limit on
the number of explosion over the entire sky ($65$, $\S~1$) is $95\%$.
This rate is much smaller than the rate of core-collapse SNe 
($r_{cc} \approx 7.5 \times 10^4~ {\rm Gpc}^{-3} {\rm yr}^{-1}$ at $<z>=0.26$ and
$\sim 1.9 \times 10^4~ {\rm Gpc}^{-3} {\rm yr}^{-1}$ at $z \approx 0$; 
Cappellaro et al. 1999, 2005) or the rate of Type Ib/c SNe which is 
$\sim 0.2$ of the total rate of core-collapse SNe. This implies that
only a small fraction, a few per cent at most, 
of the the population of core-collapse SNe release a significant fraction of
their explosion energy in the form of relativistic ejecta, as suggested by 
Berger et al. (2003) and Soderberg et al. (2005a) specifically for Type Ib/c events.
Our findings therefore rule out unified models of GRBs, XRFs, and SNe~Ib/c
as viewing-angle dependent manifestations of relativistic conical jets (e.g., 
Lamb, Donaghy, \& Graziani 2005). Alternative models, e.g., invoking 
ultra-relativistic ``cannon balls'' (Dar \& De R\'ujula 2004), are not expected to produce bright
late-time radio afterglows (Dar \& Plaga 1999), and are therefore not constrained by our
observations. Provided that the beaming fraction of GRBs and XRFs is
constrained by an independent measurement, future surveys
for orphan GRBs may be able to pin down the total rate of relativistic 
explosions, compare it to the rate of GRBs and XRFs, and probe the existence of
other types of relativistic explosions.

\subsection{Implications for Future Radio Surveys}

The follow-up observations reported here imply that the radio survey
we reported in Paper I discovered four real variable sources: an AGN,
a pulsar, a radio SN, and one unidentified, optically-faint, either Galactic 
or high-redshift source.  As discussed in great detail in $\S~4.1$ of Paper I, our
survey effectively covered $\sim1/17$ of the sky at the $6$ mJy level.
Our results therefore lead us to project that $\sim70$ real sources
would have been discovered in a similar all-sky survey. Obviously,
more sensitive surveys will discover many more such events. For
instance, assuming Euclidean space and no source evolution, we predict
$>1000$ variable sources in a similar survey with a detection limit of
$F_{\rm limit}=1$ mJy (since the number of sources is proportional to
$F_{\rm limit}^{-{3\over2}}$). Considering the fact that sources such as
AGNs have steep luminosity functions (i.e., that there are many more
faint than bright sources), this is probably a very conservative lower
limit. Additionally, our search was restricted to truly transient
sources (which are not detected at all in one of the epochs) while
the number of strongly variable sources will be much larger. 
Therefore, forthcoming surveys by instruments like the Allan
Telescope Array (ATA), and, in the more distant future, the Square 
Kilometer Array (SKA), are not only bound to discover
many such events, but will have to account for this population as a
source of systematic ``noise'' in many other types of studies.

Van Dyk et al. (2000) discuss the implications that a next-generation
radio array (in that case, SKA) will have on {\it follow-up}
observations of individual radio SNe. Let us discuss here the use of
radio surveys with the ATA as a means to {\it discover} radio SNe.

The ATA is a new radio telescope array, operated jointly by the
University of California, Berkeley (UCB) and the SETI institute, now
under construction at UCB's Hat Creek radio observatory. The first
radio dishes of the ATA are already in place, and the complete array,
consisting of 350 6.1-m radio telescopes, is expected to be completed
by the end of the decade.  The full array will be able to cover the
entire sky visible from Hat Creek ($\sim30000$ square degrees), down to
a $5\sigma$ detection limit below $1$ mJy at $1.4$ GHz, in less than
a week.  For sources with variability time scales longer than a week
(such as radio SNe), this is thus effectively a continuous survey.

Sources as luminous as VLA121550.2+130654 would be detected by the ATA
in all galaxies closer than $\sim64$ Mpc. Such a survey would
therefore produce a full census of nearby radio-bright SNe. This SN
sample will have several unique properties. Since the discovery mode
we discuss here is through an all-sky survey, the resulting sample
will not depend on the properties of the host galaxies. In particular,
this sample will be free from possible selection biases introduced by
searching for SNe only in optically bright galaxies, as done by most
of the successful optical searches responsible for discovering the
majority of nearby SNe (e.g., the KAIT search at Lick Observatory;
Filippenko 2005, and references therein). 
In addition, the background emission from
host galaxies is expected to have little effect on SN discovery in the
radio band, and thus the SN position within its host is expected to
have little effect on its discovery, as opposed to optical searches
which are less efficient near the bright nuclei of galaxies and in
edge-on spirals. Finally, as demonstrated here, a radio-selected SN
sample will be almost free from the effects of absorption by dust,
which strongly affect searches in the optical and even in the infrared
(see, e.g., Maiolino et al. 2002; Mannucci et al. 2003). 
Such a survey will still have a bias toward radio-bright SNe, and will 
need to account for the radio luminosity function of core-collapse SNe. 
Overall, though, it will provide a valuable addition to studies
based on SN optical surveys.

The real revolution is expected with the advent of the sensitive SKA. 
The accumulated experience gathered in the last few years shows that
{\it every} core-collapse SN closer than 10 Mpc, and certainly
below 5 Mpc, is detectable in the radio with the VLA (Weiler et al. 
2002; Berger et al. 2003). Put in other words, there are no "radio-quiet"
SNe among nearby events, including those SN subtypes (e.g., SNe II-P)
that have been considered to be radio-quiet in earlier literature\footnotemark.
\footnotetext{We emphasize this is true only for core-collapse events. SNe Ia, 
generally considered to result from thermonuclear explosions of white
dwarf stars, have never been detected in the radio, and may well be genuinely 
radio-quiet.}
The SKA is expected to be 100 times more sensitive than the VLA
(Cordes, Lazio, \& McLaughlin 2004), 
and so it should detect {\it every core-collapse SN} 
out to 50--100 Mpc. Thus, based on our current understanding of radio SNe,
we expect the volume-limited sample of
radio SNe discovered by an all-sky survey with the SKA will be indeed
almost unbiased, free from the effects of dust, and will not depend on the 
radio luminosity function of SNe. Such a sample will 
probe the overall SN rate, a tracer of the local star-formation rate; 
the properties of SNe as a function of host-galaxy type, color, 
morphology, and luminosity; and the distribution of
SNe within their hosts. 

Finally, in the context of our initial motivation to conduct the study
described in Paper I, all-sky variability surveys with telescopes like
the ATA and SKA will be able to discover GRB radio afterglows, both from
on-axis events (whether seen by high-energy satellites or not) and, if
these events are indeed numerous, also from relatively nearby
``orphan'' afterglows.

\subsection{Summary}

We have presented follow-up radio and optical observations of candidate
radio transients identified by comparing the FIRST and NVSS radio
surveys in search for possible orphan radio afterglows of GRBs (Paper
I). Our new observations allow us to characterize the nature of all 
of the previously discovered transients, which constitute a complete 
representative sample of radio transients down to 6 mJy a 1.4 GHz.
We conclude that none of these sources is likely
to have been an orphan GRB afterglow.

We use this fact to re-derive a lower limit on the beaming factor
$f_b^{-1}\ge62$ ($f_b^{-1}\ge20$ assuming the Frail et al. 2001 
correlation) consistent with an independent estimate by Guetta et
al. (2005; $<f_b^{-1}>=75\pm25$). We then argue that if this
consistency is to be maintained, then wider and/or deeper variability
surveys, such as those expected to be conducted with the ATA and SKA,
should detect numerous orphan afterglows; otherwise (i.e., if no
orphan afterglows are detected by such surveys) an analysis similar to
ours would result in lower limits on $f_b^{-1}$ which will greatly
exceed the Guetta et al. (2005) estimates.

We show that our survey constrains the rate of relativistic explosions
of all types, and implies that just a small fraction of core-collapse
SNe (and Type Ib/c in particular) release unconfined (e.g., conical 
jets) of relativistic ejecta, the basic ingredient in fireball models of long-soft GRBs.

Our likely detection of an optically obscured radio SN in the Virgo
spiral galaxy NGC 4216 illustrates the power of wide, sensitive, 
radio-variability surveys, such as those planned with the ATA and SKA, to
uncover a population of hidden SNe, which currently escape detection
even in the most nearby galaxies. Sensitive radio surveys may thus
provide, for the first time, a complete census of core-collapse SNe, free from
various selection biases which contaminate current compilations, such
as the tendency to monitor (and discover) SNe only in bright,
luminous galaxies, and the strong effects of host-galaxy dust
obscuration on the discovery of SNe in optical surveys. 

\section*{Acknowledgments}

We are grateful to E. Berger, M. Bietenholz, L. Blitz, P. B. Cameron, J. Cordes, A. Dar, Z. Frei,
P. Guhathakurta, J. Gunn, D. Kaplan, S. Kulkarni, A. Mahabal, P. Price, R. Sari,
and M. Sullivan for useful help, data, and advice.  A.G. acknowledges support
by NASA through Hubble Fellowship grant \#HST-HF-01158.01-A awarded by
STScI, which is operated by AURA, Inc., for NASA, under contract NAS
5-26555. A.L. and E. W. acknowledge partial support by an ISF grant for the Israeli
Center for High Energy Astrophysics. A.V.F. and W.L. are supported by National Science 
Foundation grant AST-0307894. KAIT was made possible by generous 
donations from Sun Microsystems, Inc., the Hewlett-Packard Company,
AutoScope Corporation, Lick Observatory, the NSF, the University of 
California, and the Sylvia and Jim Katzman Foundation. A.V.F. is grateful
for a Miller Research Professorship at U.C. Berkeley, during which part of
this work was completed.

\begin{deluxetable}{lllll}
\tabletypesize{\scriptsize}
\tablecaption{Flux Measurements of Candidate Radio Transients}
\tablewidth{17cm}
\tablehead{
\colhead{\#} &\colhead{Candidate} & \colhead{Data source} & \colhead{UT Date} & \colhead{1.4 GHz Flux [mJy]\tablenotemark{a}}\\
}
\startdata
1 & VLA082150.2+174616   & NVSS       & Nov. 1, 1993  & Confused with sidelobes from nearby $1.8$ Jy source\\
&                     & FIRST      & Jan. 1998 & 5.3 \\
&                     & This work  & Nov. 11, 2002 & 5.0$\pm0.8$\\
2 & VLA104848.9+551509   & NVSS       & Nov. 23, 1993  & Missed due to blending with a nearby diffuse source\\
&                     & FIRST      & Mar. 1997 & 5.2$\pm$0.16 \\
&                     & This work  & May. 18, 2002 & 5.75$\pm0.15$\\
&                     & This work  & Nov. 11, 2002 & 6.0$\pm0.3$\\
3 & VLA114355.3+221020   & NVSS       & Dec. 6, 1993  & Confused with sidelobes from nearby $2.9$ Jy source\\
&                     & FIRST      & Sep. 1998 & 5.4 \\
&                     & This work  & May. 18, 2002 & 4.6$\pm1.1$\\
&                     & This work  & Nov. 11, 2002 & 8.4$\pm1.8$\\
4 & VLA121550.2+130654   & NVSS       & Feb. 27, 1995 & Undetected ($\le1$)\\
&                     & FIRST      & Dec. 1999 & 8.6$\pm0.2$\\
&                     & This work  & May. 18, 2002 & 15.7$\pm0.3$\\
&                     & This work  & Nov. 11, 2002 & 14.6$\pm0.3$\\
&                     & This work  & Jun. 27, 2003 & 12.6$\pm0.2$\\
&                     & This work  & Mar. 06, 2004 & 13.2$\pm0.2$\\
5 & VLA122532.6+122501   & NVSS       & Feb. 27, 1995  & $1\pm0.9$\\
&                     & FIRST      & Apr. 2001 & $6.45\pm0.15$\\
&                     & This work  & Nov. 11, 2002 & $6.34\pm0.6$\tablenotemark{b}\\
6 & VLA130713.5-052709   & NVSS       & Feb. 27, 1995  & $0.1\pm0.4$\\
&                     & FIRST      & Apr. 2001 & $5.3$\tablenotemark{c}\\
&                     & This work  & Nov. 11, 2002 & $-0.09\pm0.16$\tablenotemark{d}\\
7 & VLA152248.7+542644   & NVSS       & Nov. 23, 1993  & $1.8\pm0.5$\\
&                     & FIRST      & May. 1997 & $6.1$\\
&                     & This work  & Nov. 11, 2002 & $7.7\pm0.5$\tablenotemark{e}\\
8 & VLA165203.1+265140   & NVSS       & Apr. 16, 1995  & $-0.1\pm0.5$\\
&                     & FIRST      & Dec. 17, 1995  & $5.3\pm0.15$\\
&                     & This work  & Oct. 18, 2002 & $0.7\pm0.2$\\
&                     & This work  & Nov. 11, 2002 & $0.5\pm0.2$\tablenotemark{f}\\
9& VLA172059.9+385227   & NVSS       & Apr. 19, 1995  & $-0.9\pm0.5$\\
&                     & FIRST      & Aug. 7,  1994  & $9.4\pm0.2$\tablenotemark{g}\\
&                     & This work  & Oct. 18, 2002 & Undetected ($-0.07\pm0.1$)\\
&                     & This work  & Nov. 11, 2002 & Undetected ($0.01\pm0.1$)\tablenotemark{h}\\
\enddata
\tablenotetext{a}{We have remeasured the flux in archival FIRST data, and report here our
revised measurements, which are slightly different (typically by less than $10\%$) from those
reported in the FIRST catalog and used in Paper I.}
\tablenotetext{b}{An X-band flux of $4.76\pm0.05$ mJy measured on the same date indicates
a flat spectrum with power-law index $\alpha=-0.16$ (where $F_\nu \propto \nu^{\alpha}$).}
\tablenotetext{c}{Nearby unusual negative-flux features and unexplained elevated noise 
levels cast doubt on the reality of this detection.}
\tablenotetext{d}{Undetected also in contemporaneous X-band measurements ($-0.02\pm0.05$ mJy).}
\tablenotetext{e}{Measurements of this source are compromised by a combination of a nearby
source with similar flux which is not well resolved by NVSS, and elevated noise levels from
a $1.3$ Jy source just $13'$ away. Comparing our best reduction of the higher-resolution
data (FIRST vs. our own observations) we find that this source is most likely constant.}
\tablenotetext{f}{Known radio pulsar PSR J1652+2651.}
\tablenotetext{g}{In this case FIRST data were obtained prior to NVSS data.}
\tablenotetext{h}{Similar limits are obtained at X-band. Other sources in the VLA field are constant.}
\end{deluxetable}

\clearpage
\begin{deluxetable}{lcrcrr}
\tabcolsep0.1in\footnotesize
\tablewidth{\hsize}
\tablecaption{Radio Observations of VLA\,121550.2+130654 \label{tab:data}}
\tablehead {
\colhead {Epoch}      &
\colhead {Frequency} &
\colhead {Flux Density} &
\colhead {RMS} &
\colhead {Array} \\
\colhead {(UT)}      &
\colhead {(GHz)} &
\colhead {(mJy)} &
\colhead {(mJy)} &
\colhead {Config.} &
}
\startdata
1990 Jan. 07 &  1.43 &  $<$0.15   & \nodata  &  D \\
1995 Feb. 27 &  1.43 &  $<$1.0    & \nodata  &  D (NVSS)   \\
1999 Dec.    &  1.43 &  8.6       & 0.2  & B (FIRST) \\
2002 May  19 &  1.43 &  15.7      & 0.3  & BnA \\
2002 May  19 &  8.46 &  2.60      & 0.06 & BnA \\
2002 Nov. 11 &  1.43 &  14.6      & 0.3  & C \\
2002 Nov. 11 &  8.46 &  2.93      & 0.04 & C \\
2003 Jun. 27 &  1.43 &  12.6      & 0.2  & A  \\
2003 Jun. 30 &  4.86 &  4.99      & 0.06 & A \\
2003 Jun. 30 &  8.46 &  2.76      & 0.05 & A \\
2003 Aug. 25 &  8.46 &  2.88      & 0.07 & A \\
2004 Mar. 06 &  1.43 &  13.2      & 0.2  & C \\
2004 Mar. 06 &  8.46 &  2.89      & 0.04 & C       \\
2004 May. 02 &  1.43 &  9.63      & 0.4  & VLBA    \\
2004 May. 02 &  8.46 &  2.5       & 0.5  & VLBA    \\
\enddata
\tablecomments{The columns (left to right) are (1) the UT date of each
   observation, (2) observing frequency, (3) flux density at the
   position of the radio transient, (4) RMS noise calculated from
each image, and (5) VLA array configuration. }
\end{deluxetable}

\begin{deluxetable}{llll}
\tablecaption{Optical Observations of the Location of VLA172059.9+385227}
\tablewidth{17cm}
\tablehead{
\colhead{UT Date} & \colhead{Telescope} & \colhead{Camera} & \colhead{Exposure times and filters}\\
}
\startdata

2003 Aug. 8          & Wise 1m    & Tektronics 1024$^2$ CCD & 150 s $V$, 150 s $R$, 150 s $I$\\
2003 Mar. 7          & P200       & LFC                     & 600 s $R$ \\
2003 Jun. 28         & P200       & LFC                     & 1200 s $R$ \\
2004 Apr. 22         & Keck I     & LRIS R+B                & 300 s $g$, 300 s $I$ \\
2004 Nov. 4          & P60        & 2048$^2$ CCD            & 900 s $g$, 900 s $R$, 900 s $I$ \\
\enddata
\end{deluxetable}



\begin{thebibliography}{}

\bibitem{} Bartel, N., \& Bietenholz, M.~F.\ 2003, \apj, 591, 301 
\bibitem{} Berger, E., Kulkarni, S. R., \& Frail, D. A. 2003, \apj 590, 379
\bibitem{} Berger, E., Kulkarni, S.~R., Frail, D.~A., \& Soderberg, A.~M.\ 
   2003, \apj, 599, 408 
\bibitem{} Berger, E., et al.\ 2002, \apj, 581, 981 
\bibitem{} Bloom, J.~S., Frail, D.~A., \& Kulkarni, S.~R.\ 2003, \apj, 594, 674 
\bibitem{} Bower, G.~C., Roberts, D.~A., Yusef-Zadeh, F., Backer, D.~C.,
  Cotton, W.~D., Goss, W.~M., Lang, C.~C., \& Lithwick, Y.\ 2005, ApJ, 
  in press, ArXiv Astrophysics e-prints, astro-ph/0507221
\bibitem{} Cappellaro, E., Evans, R., \& Turatto, M.\ 1999, \aap, 351, 459 
\bibitem{} Cappellaro, E., et al.\ 2005, \aap, 430, 83
\bibitem{} Carilli, C.,~\& Rawlings, S.\ 2004, ArXiv Astrophysics e-prints, 
   astro-ph/0409274
\bibitem{} Claver, C.~F., et al.\ 2004, \procspie, 5489, 705
\bibitem{} Condon, J.~J., Cotton, W.~D., Greisen, E.~W., Yin, Q.~F., Perley, 
  R.~A., Taylor, G.~B., \& Broderick, J.~J.\ 1998, \aj, 115, 1693
\bibitem{} Cordes, J.~M., Lazio, T.~J.~W., \& McLaughlin, M.~A.\ 2004, ArXiv 
  Astrophysics e-prints, astro-ph/0410045 
\bibitem{} Dar, A., \& Plaga, R.\ 1999, \aap, 349, 259 
\bibitem{} Dar, A. \& De R\'ujula, A. 2004 Phys. Rep. 405, 203
\bibitem{} Djorgovski, S.~G., et al.\ 2004, American Astronomical Society Meeting, 204, $\#$75.03
\bibitem{} Filippenko, A. V. 2005, in The Fate of the Most Massive
    Stars, ed. R. Humphreys and K. Stanek (San Francisco: ASP), 34
\bibitem{} Filippenko, A. V., Li, W., Treffers, R. R., \& Modjaz, M. 2001, in 
    Small-Telescope Astronomy on Global Scales, ed. W.-P. Chen, C. Lemme, \&
    B. Paczy\'{n}ski (San Francisco: ASP), 121
\bibitem{} Foley, R.~J.,~et al.\ 2003, \pasp, 115, 1220 
\bibitem{} Frail, D. A., Waxman, E., \& Kulkarni, S. R. 2000, \apj, 537, 191
\bibitem{} Frail, D. A., et al. 2001, \apj, 562, L55
\bibitem{} Freedman, D. L., \& Waxman, E. 2001, ApJ, 547, 922
\bibitem{} Gal-Yam, A., Maoz, D., Guhathakurta, P., \& Filippenko, A.~V.\ 2003, 
  \aj, 125, 1087 
\bibitem{} Gal-Yam, A., Ofek, E.~O., Filippenko, A.~V., Chornock, R., \& Li, 
   W.\ 2002, \pasp, 114, 587
\bibitem{} Gal-Yam, A., Ofek, E.~O., \& Shemmer, O.\ 2002, \mnras, 332, L73 
\bibitem{} Gal-Yam, A., et al.\ 2004, \apjl, 609, L59 
\bibitem{} Graham, J.~A., et al.\ 1999, \apj, 516, 626 
\bibitem{} Granot, J., \& Loeb, A.\ 2003, \apjl, 593, L81 
\bibitem{} Guetta, D., Piran, T., \& Waxman, E.\ 2005, \apj, 619, 412 
\bibitem{} Heise, J., in't Zand, J., Kippen, R.~M., \& Woods, P.~M.\ 2001, 
   in ``Gamma-ray Bursts in the Afterglow Era'', ed. E. Costa, F. Frontera, 
   \& J. Hjorth, (Berlin Heidelberg: Springer), 16 
\bibitem{} Hurley, K., et al.\ 2002, \apj, 567, 447 
\bibitem{} Hyman, S.~D., Lazio, T.~J.~W., Kassim, N.~E., Ray, P.~S., 
  Markwardt, C.~B., \& Yusef-Zadeh, F.\ 2005, \nat, 434, 50 
\bibitem{} Jaunsen, A.~O., et al.\ 2003, \aap, 402, 125 
\bibitem{} Kaiser, N.\ 2004, \procspie, 5489, 11 
\bibitem{} Kulkarni, S.~R., \& Phinney, E.~S.\ 2005, \nat, 434, 28
\bibitem{} Kulkarni, S.~R., et al.\ 1998, \nat, 395, 663
\bibitem{} Lamb, D.~Q., Donaghy, T.~Q., \& Graziani, C.\ 2005, \apj, 620, 355
\bibitem{} Landolt, A.~U.\ 1983, \aj, 88, 439
\bibitem{} Lee, B.~C., et al.\ 2003, Gamma-Ray Burst and Afterglow Astronomy 
  2001: A Workshop Celebrating the First Year of the HETE Mission, Ed. G. R. Ricker
  and R. K. Vanderspek (New York: AIP), p. 349 
\bibitem{} Levinson, A., Ofek, E.~O., Waxman, E., \& Gal-Yam, A.\ 2002, 
   \apj, 576, 923 (Paper I)
\bibitem{} Li, W., et al. 2000, in Cosmic Explosions, ed. S. S. Holt \& 
  W. W. Zhang (New York: AIP), 103
\bibitem{} Linder, E.~V., et al.\ 2005, ArXiv Astrophysics e-prints, astro-ph/0406186 
\bibitem{} Mahabal, A., et al. 2005, in ``Wide Field Imaging from Space,'' 
   ed. T. McKay, A. Fruchter, \& E. Linder, New Astronomy Reviews, astro-ph/0408035 
\bibitem{} Maiolino, R., Vanzi, L., Mannucci, F., Cresci, G., Ghinassi, F., \& 
   Della Valle, M.\ 2002, \aap, 389, 84 
\bibitem{} Mannucci, F., et al.\ 2003, \aap, 401, 519 
\bibitem{} Monet, D.~G., et al.\ 2003, \aj, 125, 984
\bibitem{} Read, A.~M., Saxton, R.~D., Esquej, M.~P., Freyberg, M.~J., \& Altieri, B.\ 2005, ArXiv 
Astrophysics e-prints, arXiv:astro-ph/0506380 
\bibitem{} Rhoads, J.~E.\ 1997, \apjl, 487, L1 
\bibitem{} Rhoads, J.~E.\ 2003, \apj, 591, 1097 
\bibitem{} Soderberg, A.~M., Kulkarni, S.~R., Berger, E., Chevalier, R.~A.,
   Frail, D.~A., Fox, D.~B., \& 
Walker, R.~C.\ 2005, ArXiv Astrophysics e-prints, astro-ph/0410163 
\bibitem{} Soderberg, A.~M., Nakar, E., \& Kulkarni, S.~R.\ 2005a, ApJ, 
  submitted, ArXiv Astrophysics e-prints, arXiv:astro-ph/0507147
\bibitem{} Soderberg, A.~M., et al.\ 2005b, \apj, 627, 877
\bibitem{} Stocke, J.~T., Morris, S.~L., Weymann, R.~J., \& Foltz, C.~B.\ 1992, 
   \apj, 396, 487 
\bibitem{} Sullivan, M.,~et al.\ 2004, ArXiv Astrophysics e-prints, astro-ph/0410594 
\bibitem{} Turolla, R., Possenti, A., \& Treves, A.\ 2005, \apjl, 628, L49
\bibitem{} Van Dyk, S.~D., Weiler, K.~W., Montes, M.~J., Sramek, R.~A., \&
  Panagia, N.\ 2000, in Perspectives on Radio Astronomy: Science with Large 
  Antenna Arrays, ed. M.~P.~van Haarlem.~Published by ASTRON., p. 241
\bibitem{} Vreeswijk, P.~M., et al.\ 2001, \apj, 546, 672 
\bibitem{} Weiler, K.~W., Panagia, N., Montes, M.~J., \& Sramek, R.~A.\ 2002, 
  \araa, 40, 387 
\bibitem{} Weiler, K.~W., Van Dyk, S.~D., Discenna, J.~L., Panagia, N., \&
  Sramek, R.~A.\ 1991, \apj, 380, 161 
\bibitem{} White, R.~L., Becker, R.~H., Helfand, D.~J., \& Gregg, M.~D.\ 1997, 
   \apj, 475, 479 
\bibitem{} Young, C.~K.,~\& Currie, M.~J.\ 1998, \aaps, 127, 367 

\end{thebibliography}
\end{document}